\documentclass[10pt,a4paper,twoside]{article}
\usepackage{epsfig}
\usepackage{baltlat5}
\usepackage{wrapfig}
\begin{document}
\ \
\vspace{-0.5mm}

\setcounter{page}{1}
\vspace{-2mm}

\titlehead{Baltic Astronomy}% , vol.\ts 14, xxx--xxx, 2005.}

\titleb{BINARY CENTRAL STARS OF PLANETARY NEBULAE}

\begin{authorl}
\authorb{Albert A. Zijlstra}{}
 \end{authorl}
\begin{addressl}
\addressb{}{University of Manchester,
   School of Physics \&\ Astronomy,
   P.O. Box 88, Manchester M60 1QD, UK}

\end{addressl}

%\submitb{not yet}
\begin{summary}
This paper reviews our knowledge on binary central stars of planetary nebulae
and presents some personal opinions regarding their evolution.  Three types of
interactions are distinguished: type I, where the binary companion induces the
mass loss; type II, where it shapes the mass loss but does not enhance it; type
III, where a wide orbit causes the centre of mass to move, leading to a spiral
embedded in the wind.  Surveys for binary central stars are discussed, and the
separations are compared to the distribution for binary post-AGB stars. The
effect of close binary evolution on nebular morphology is discussed.
Post-common-envelope binaries are surrounded by thin, expanding disks,
expelled in the orbital plane.  Wider binaries give rise to much thicker
expanding torii. Type~I binary evolution predicts a wide distribution of
masses of central stars, skewed towards low masses.  Comparison with observed
mass distributions suggests that this is unlikely to be the only channel
leading to the formation of a planetary nebula.  A new sample of compact Bulge
nebulae shows about 40\%\ of nebulae with binary-induced morphologies.

\end{summary}

\resthead{Binary central stars}{A.A. Zijlstra}

\sectionb{1}{INTRODUCTION}

The traditional view of stellar evolution on the Asymptotic Giant Branch (AGB)
and beyond is that of a single star. An AGB star consists of an inert
carbon-oxygen core, surrounded by a nuclear-burning shell. This shell
alternates between phases of hydrogen and of helium burning, punctuated by
helium shell flashes, the thermal pulses. The nuclear burning region in turn
is surrounded by the convective envelope.  Radial pulsations increase in
amplitude and period whilst the star ascends the AGB. During the Mira phase the
pulsation periods are 150--400 days and the bolometric amplitude a magnitude or
more. (The visual amplitude can exceed 8 magnitudes, amplified by variable
molecular bands.) 

The strong pulsations lead to the formation of an extended atmosphere. In the
outer regions dust condenses: radiation pressure on the dust now drives a
highly efficient stellar mass loss. The mass-loss rate greatly exceeds the
nuclear-burning rate, and depletes the envelope. Once the envelope mass is
reduced to $M_e \sim 0.02\,\rm M_\odot$, the photosphere collapses and the
effective temperature increases.  The wind reduces or ceases: the star is now
surrounded by a detached, expanding envelope. Once the star is hot enough to
ionize the ejecta, a planetary nebula (PN) forms.

Although there is strong observational evidence for this scenario (e.g. Habing
1996), doubts have been expressed. The ejecta commonly are non-spherically,
and there is no clear mechanism for a single AGB star to eject a strongly
a-spherical nebula (Soker 1998). The efficiency of the dust-driven wind has
also been questioned (Woitke 2006). 

Binary companions can affect the mass loss in several ways. Close companions
will evolve through a common envelope phase, leading to rapid envelope
ejection: this may be called type I.  More distant companions interact with
and shape the wind but do not enhance the mass loss rate: type II. Very wide
binaries cause the centre of mass to shift, leading to a spiral embedded in
the wind (Mauron \&\ Huggins 2006), but have no other effect (type III).
Thus, type-I is binary-induced mass loss, type-II binary-shaped mass loss,
and type-III orbital shaping of the mass loss.

\sectionb{2}{SURVEYS}

  Detections of binary companions are done in four different ways.  First,
direct CCD imaging reveals distant companions. Second, spectroscopy reveals
cool (non-ionizing) components to the stellar spectrum.  Third, photometric
monitoring shows brightness fluctuations due to rotation of the heated surface
of a close companion, or in a few cases eclipses. Fourth, radial velocity
variations trace the orbital motion. There are some caveats to these
methods. Distant companions may be line-of-sight coincidences. Eclipses can be
due to orbiting dust clouds, as in the case of NGC 2346 (e.g. Roth et
al. 1984). Radial velocities can be affected by wind variations: only strict
periodicity can be taken as evidence for a binary companion, but it has proven
difficult to acquire the temporal sampling required for this.

The main survey for spatially resolved binaries is that of Ciardullo et
al. (1999) using an HST snapshot survey of 113 nearby systems. They find 19
possible companions, of which approximately 6 are expected to be due to
confusion.  The target selection criteria for the Ciardullo et al. survey
included suspected binarity, so that converting the detection rate to a binary
fraction has some uncertainty.  Roughly 10\%\ of PNe are found to have distant
($10^2$--$10^4$\,AU) companions. The detected companions are mostly
main-sequence stars. This is expected for reasons of sensitivity, as white
dwarf companions will be faint and red giant companions short-lived.  More
recently, Benetti et al. (2003) detected a companion in NGC\,6818 but this
requires confirmation. No other surveys have been done. Adaptive optics using
the central star may now be competitive with the HST observations.

More is known on non-resolved binaries.  De Marco (2006) presents a list of 25
close-binary central stars.  To this may be added Me1-1 (Shen et al. 2004)
which has a K3-4 giant companion but the orbital parameters are not known and
it may be a symbiotic star. On the other hand, the evidence for the claimed
companion of NGC\,6302 is unconvincing and this object should be removed for
now.  Binary periods range from hours to days. The longest known period is 68
days for Sh\,2-71 (Jurcik 1993).  The central star of LoTr5 is believed to be
a triple; one photometric period of 5.9 days is believed to be the rotation
period (Strassmeier et al. 1997). A63 is a known triple system, with an
11-hour eclipsing component and a third star 2.8 arcsec away (Ciardullo et
al. 1999).  Among the confirmed close binaries, four objects are eclipsing.

Sensitivity to companions at distances of 1--10 AU is poor: only radial
velocity surveys are sensitive to such companions. For A35, a resolved
companion at 18\,AU was found by Gatti et al. (1998). Kinematic evidence for
Hu2-1 suggests a binary with separation of 9--27\,AU (Miranda et al. 2001) but
the evidence is indirect.

Selection effects need to be considered. Detection of distant companions is
easiest for faint central stars and large, faded nebulae. On the other hand,
detection of close binaries requires a relatively bright star with a faint
nebula. This is found in systems where the star evolves much slower than the
nebula, i.e. the nebula has had time to expand but the star is still
relatively cool. This favours low-mass central stars.  The requirement for a
bright central star may also directly select binaries if the visual brightness
of the companion is similar to or exceeds that of the central star.

\vbox{
\centerline{\psfig{figure=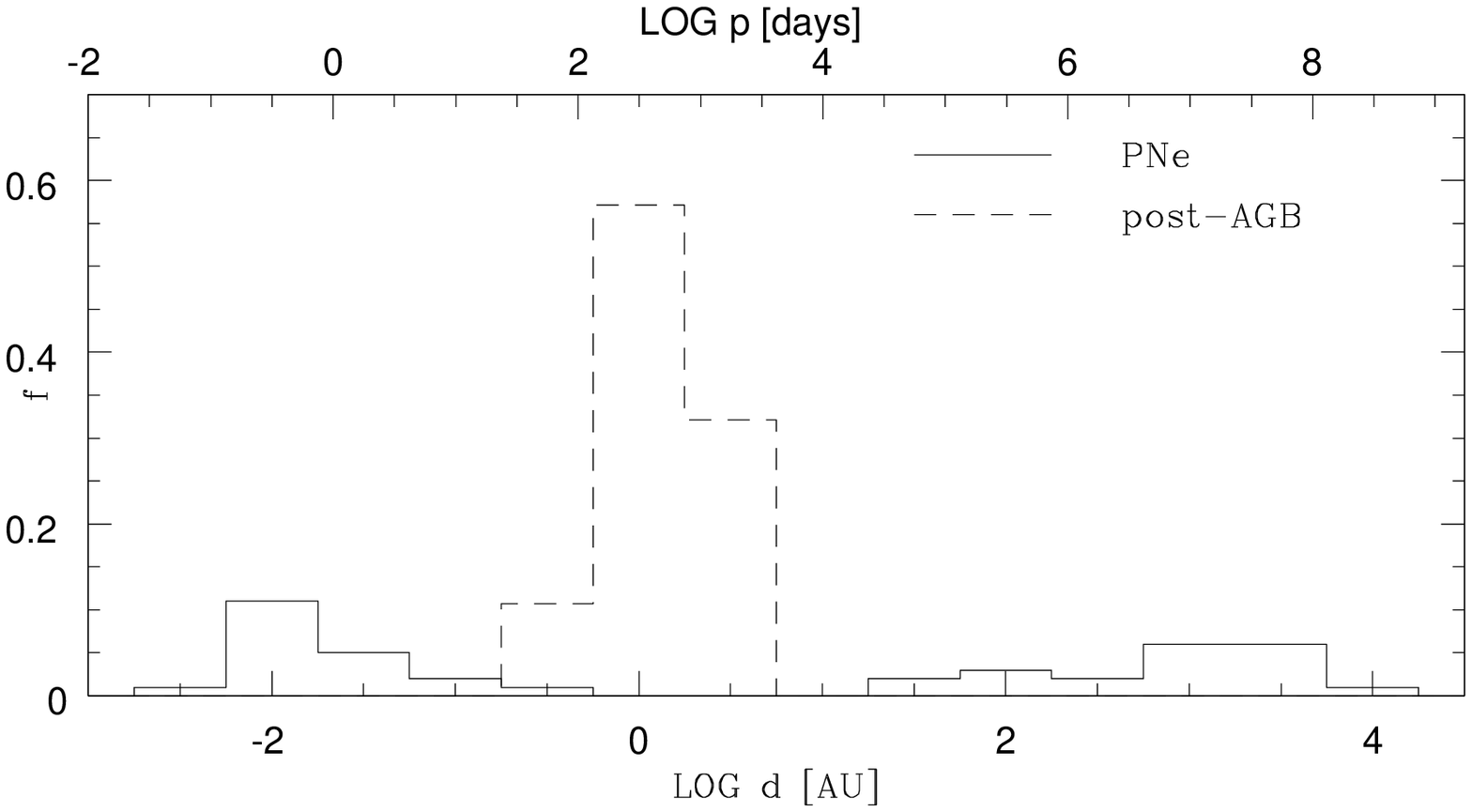,width=120truemm,angle=0,clip=}}
\vspace{-5mm}
\captionc{1}{Observed binary separations as fraction of the population, for
planetary nebulae and post-AGB stars.{\it Based on data kindly provided by
van Winckel.} }
}
\vspace{3mm}

Surveys for post-AGB stars have uncovered a number of stars located on the HR
diagram between the AGB and the PNe. These include optically bright stars with
circumstellar dust. Surveys have shown that these are invariably binaries,
with periods of typically 300--1200 days (van Winckel 2003). Fig. 1 shows the
orbital separations (assuming a total mass of 1.4\,M$_\odot$) and periods,
compared to those of the binary PNe. All are converted to a population
fraction, where for PNe I assume that all known binaries come from the $\sim
100$ PNe which have stars bright enough to be easily observable.  The lack of
overlap is remarkable.

On the short-period side, the PNe systems must have gone through a
common-envelope phase. The post-AGB stars have avoided this, as indicated by
the ellipticicty of the orbits. Thus, these are distinct populations.  Very
wide binaries are unobservable among post-AGB stars because the stars are too
bright. However, one can expect that if a large fraction of PNe central stars
were binaries with orbital separations similar to the post-AGB stars, this
would have been discovered. It seems probable that the binary post-AGB stars
do not evolve into 'typical' PNe.

\sectionb{3}{MORPHOLOGIES}

The standard morphologies of planetary nebulae are: round, elliptical, bipolar
(or butterfly), and irregular (Balick \&\ Frank 2002). Binary interactions are
the main proposed origin of the non-spherical structures, although different
precise mechanisms have been suggested. The binary companion acts as a source
of angular momentum, either to the stellar ejecta or, in the case of a common
envelope, directly to the star.

The most pronounced morphology is that of the butterfly shape. One might
predict that PNe around close binary stars would show this shape as they
experience the most efficient transfer of angular momentum to the ejecta.
However, observations do not fully confirm this. Of the three systems with
possible CV nuclei (HFG1, A65, K1-2) none are bipolar (Walsh \&\ Walton 1996).
Only one butterfly nebula has a confirmed binary nucleus (NGC 2346): it has a
period of 16 days, among the longer periods known.

Planetary nebulae around closer binaries do show deviations from spherical
symmetry (Bond \&\ Livio 1991): about half appear bipolar or elliptical, and
one shows a jet-like structure. One is perfectly round, but Bond \&\ Livio
argue this object (Sp1) is likely seen pole-on. This was confirmed by Mitchell
(priv. comm.) from kinematics of the nebula.  These objects lack the
characteristic double shell morphology of normal PNe. The last point suggests
that they evolve differently from other, 'normal' PNe.

Overall, rather irregular structures appear to be the norm for the closest
binaries, as for example shown in Fig.  2 (from Pollacco \&\ Bell 1997). An
interesting case is A63, which shows an expanding, edge-on torus (shown in
Fig. 3) and two polar blobs several arcminutes from the star. They trace a
tightly collimated flow.  Mitchell et al. (2006) show that the polar
ejection is almost in the plane of the sky. The central star is known to be
eclipsing. This is strong evidence that the torus is expelled in the orbital
plane and that the collimated outflow is perpendicular to the orbital plane.
This implies that the nebula was expelled by type-I binary interaction.

\vspace{3mm}
\vbox{
\centerline{\psfig{figure=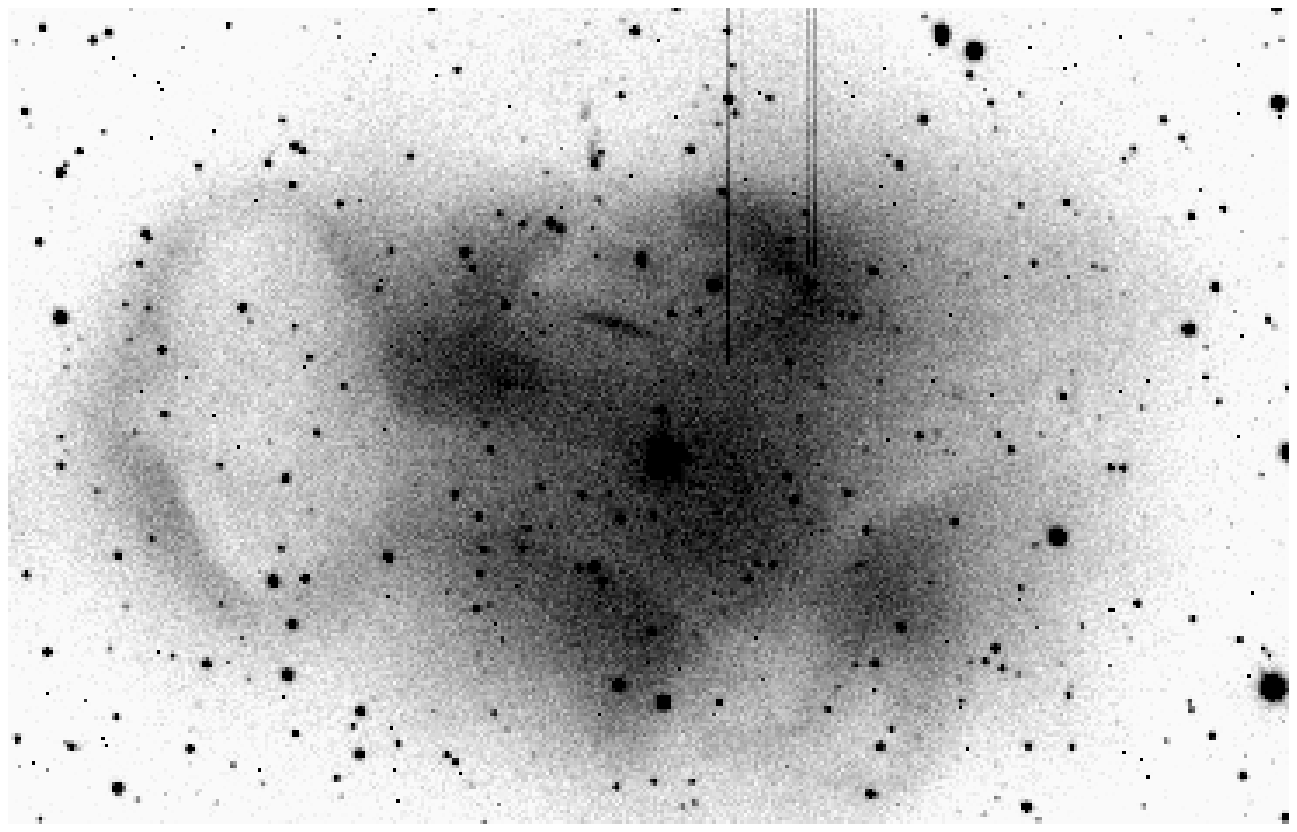,width=100truemm,angle=0,clip=}}
\vspace{-5mm}
\captionc{2}{The non-eclipsing PN Ds1 (0.45 days). }
}
\vspace{3mm}

\vbox{
\centerline{\psfig{figure=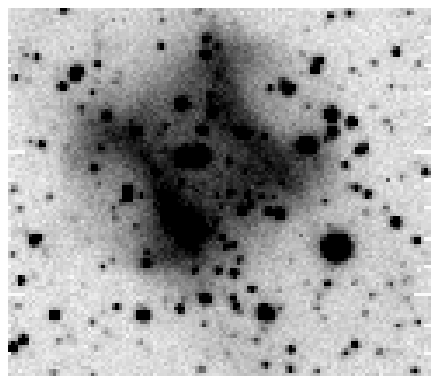,width=50truemm,angle=0,clip=}}
\vspace{-5mm}
\captionc{3}{The eclipsing PN A63 (0.45 days). }
}
\vspace{3mm}

An interesting suggestion is that PNe with the closest (hours to days) binary
stars show thin, expanding rings, while somewhat longer periods (tens of days)
give thick expanding torii and lead to butterfly nebulae.  The thin rings
(seen in objects such as SuWt\,1, A\,61 and WeBo\,1: e.g. Bond 2000) contain
less mass but are dynamically cold and may contain more angular momentum.  The
torii are more massive but contain limited angular momentum.  These two
categories may trace different types of interaction. The thin ring may contain
material lost through the outer Langrangian point.

It appears that  known post-common-envelope systems have identifiable
morpologies which are not particularly common among planetary nebulae. Whether
this holds for all such systems remains to be seen.

\sectionb{4}{POPULATION STUDIES}

Corradi et al. (1995) find that 15 per cent of Galactic PNe are bipolar. The
bipolars have a smaller scale height in the Galaxy than other types,
indicating on average a higher progenitor mass. If the link between bipolarity
and binaries holds, this indicates that 15 per cent of PNe progenitors have
companions close enough to affect the mass loss.  An unpublished survey of
compact Bulge PNe carried out with HST shows 27 per cent to be bipolar; an
additional 13 per cent show morphologies which may be post common envelope
systems: multi-polar, spiral (one object), or thing ring structures.  Adding
the bipolar/butterfly systems would give a maximum fraction of type-I
interaction of order of 40 per cent.  New nearby bipolar nebulae are still
being discovered (Frew et al. 2006) and the percentage among Galactic disk PNe
could increase.

Detailed population studies are presented by de Marco \&\ Moe (2006).  About
half of stellar systems are known to be multiple.  Of the multiple systems,
about 25 per cent have wide separations in the range traced by Ciardullo et
al., and a similar number are so close that interaction is expected during the
post-main sequence evolution.  Of the latter, about a third will form a common
envelope on the first giant branch, and will never reach the AGB, unless a
merger occurs first. They are unlikely to form PNe. The remaining fraction is
not inconsistent with the  fraction of post-common envelope systems
found in the Bulge, and the fraction of detected binary systems shown in
Fig.~1.

However, de Marco \&\ Moe (2006) arrive at a different conclusion, and argue
that only close binary interactions lead to PN, i.e. the fractions above
become 100 per cent. This requires that single-star mass loss is insufficient
to create a dense PN.  Soker \&\ Subag (2005) predict that single stars give
rise to fainter PNe, which are under-represented in existing samples.

The easiest way to distinguish between type-I binary evolution models and all
other models is to test the predicted final masses of the stars.  To obtain
observed mass distributions, Gesicki \&\ Zijlstra (2000) use diagrams of
dynamical age versus stellar tempature to infer the rate of temperature
increase of the star, which is a strong function of mass of the
(pre-white-dwarf) central star. These diagrams measure the stellar masses,
much more accurately than can be done with photometric or spectroscopic
methods.  They a find a narrow distribution, between 0.57 and 0,65\,M$_\odot$
(Fig. 4, left panel). (Note that the method used is retricted to regular PNe
and that therefore the sample excludes bipolar nebulae.)  The observed
distribution is consistent with expectations: for lower masses, stars evolve
so slowly that the nebula disperses before the onset of ionization. At higher
masses the evolution is so fast that there is little chance of catching an
object in this phase.

\vspace{5mm}
\vbox{
\centerline{\psfig{figure=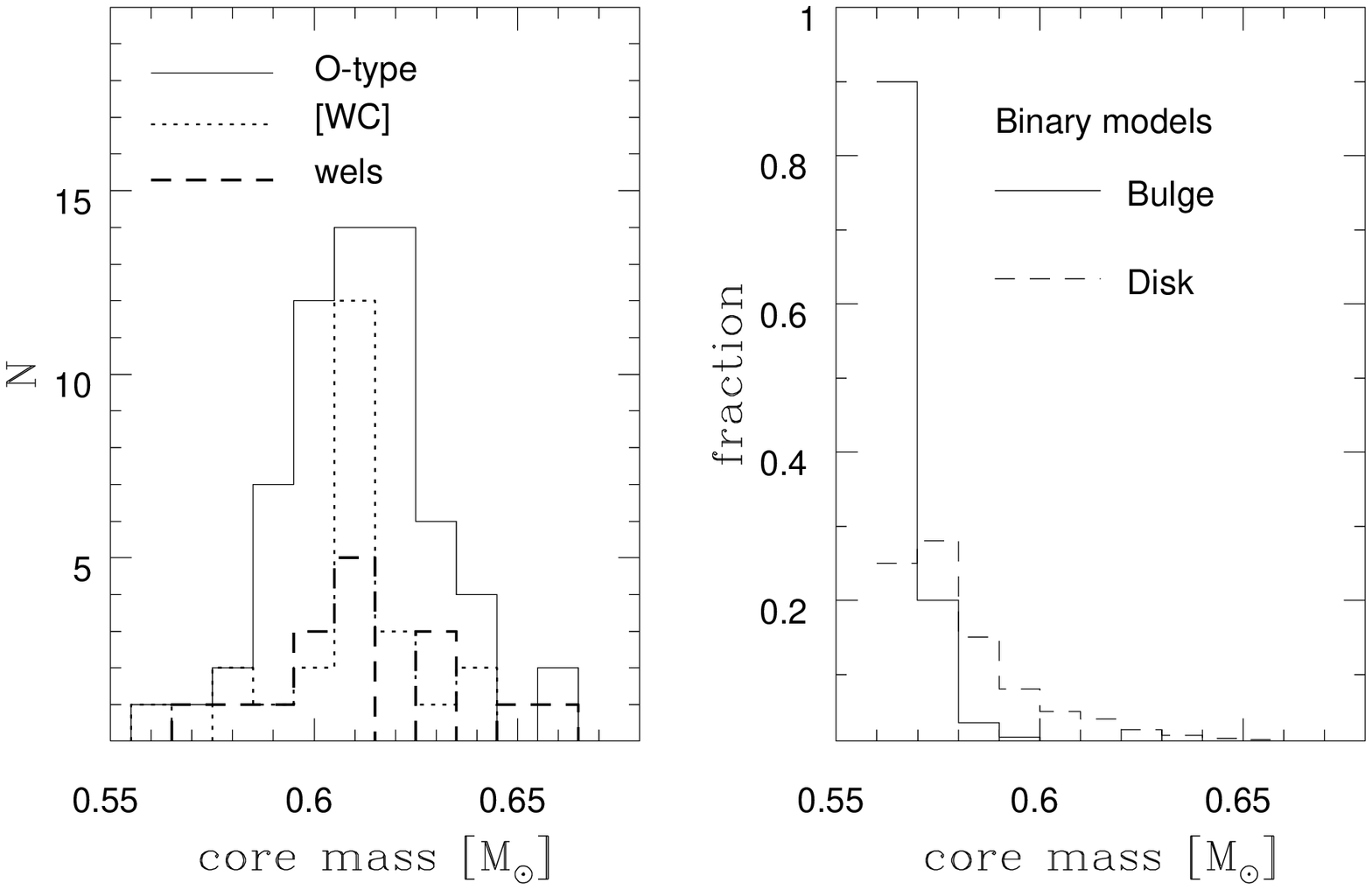,width=60truemm,angle=0,clip=}}
\vspace{-5mm}
\captionc{4}{Left: Mass distribution of central stars of PNe (from Gesicki et
  al. 2006). Right: predictions from type-I binary models (from de Marco \&\
  Moe 2006).}
}
\vspace{3mm}

The narrow mass range fits in well with our knowledge of AGB evolution: the
mass loss is linked to stellar parameters such as the core mass, and the AGB
evolution is terminated at a fixed point where the mass-loss rate exceeds the
nuclear burning rate.  But common envelope evolution does not predict such a
narrow range. The end point of evolution depends on the mass and orbital
parameters, and a wider range of resulting masses may be expected, biased
towards lower final masses than obtained from single-star AGB
evolution. The right panel of Fig. 4 shows the binary-model predictions
from Marco \&\ Moe (2006) (their Fig. 11). As expected, it is strongly peaked
towards lower masses, and differs significantly from observed distributions.

\sectionb{5}{EVOLUTION}

 The stellar mass distribution of Fig.~4 indicates that common envelope
evolution is not the dominant evolutionary branch on the AGB. However, it can
account for a larger fraction of the lower-mass PNe central stars.

The observations now suggest several distinct routes to the post-AGB region of
the HR diagram. Single-star evolution leads to normal PNe. Type II
interactions also evolve on this track. Type I evolution early on the AGB
leads to a low mass remnant, which evolves too slowly to become a PN. These
may be the binary post-AGB stars discussed above. Finally, type-I evolution
during the thermal-pulsing AGB leads to a remnant which evolves sufficiently
fast to become a PN, leading to the 'thin ring' or 'thick torus' nebulae.

\sectionb{6}{CONCLUSIONS}

The established models for AGB/post-AGB/PNe stellar evolution are in terms of
single-star evolution. But the nebular morphologies suggest that binary
interactions do have an important role to play.  The work by de Marco \&\ Moe
(2006) has invigorated the field. Whether binary interactions are as dominant
as argued by them remains to be proven, but there is a strong case that the
binary channel is a significant one for post-AGB evolution.

Observationally, we can distinguish three types of interactions: binary-induced
mass loss (type I), binary-shaped mass loss (type II) and shaping by orbital
motion (type III).  From present samples, it appears that type-I interactions
lead to identifiable morphologies, characterized by expanding rings and
collimated outflows for the closest binaries, and thick expanding torii
tracing somewhat wider binaries.  The physical parameters determining the
final result are not known, but bipolar nebulae tend to have higher mass
progenitors.  Type-II interactions are probably very common, as shown by the
companion to Mira itself and the possible shaping of its wind (e.g. Josselin
et al. 2000). These are therefore a leading contender for the main
shaping mechanism for planetary nebulae.

Surveys for binary companions have not yet been completed. To spatially
resolve systems, ground-based adaptive optics may proof to be competitive with
HST. Binaries of separations of order 1 AU are missing among the PN samples:
velocity monitoring should be attempted to find this missing link.

\vskip5mm

\References

\refb Balick B., Frank A., 2002, ARA\&A, 40, 439

\refb Benetti S., Cappellaro E., Ragazzoni R., Sabbadin F., Turatto M.,
2003, A\&A, 400, 161

\refb Bond H.E., 2000, Binary Stars in Planetary Nebulae, 
Encyclopedia of Astronomy and Astrophysics, Edited by Paul Murdin
 (Bristol: Institute of Physics Publishing)

\refb Bond H.E., Livio M., 1990, ApJ, 355, 568

\refb Ciardullo R., Bond H.\,E., Sipior M.\,S., Fullton L.\,K.,
 Zhang C.-Y., Schaefer K.\,G., 1999, AJ, 118 488

\refb Corradi R.L.M., Schwarz H.E., 1995, A\&A, 293, 871

\refb de Marco O., 2006, IAU Symposium 234, Planetary
Nebulae. astro-ph/0605626

\refb de Marco O., Moe M., 2006, ApJ, 

\refb Feibelman W.\,A., 2001, ApJ, 550, 785

\refb Frew D.J., Parker Q.A., Russeil, D., 2006, MNRAS, in press

\refb Gatti A.A., Drew J.E., Oudmaijer R.D., Marsh T.R., Lynas-Gray
A.E., 19998, MNRAS, 301, L33

\refb  Gesicki K., Zijlstra A.A., 2000, A\&A, 358, 1058

\refb Gesicki K., Zijlstra A.A., Acker A., G\'orny S.K.,
 Gozdziewski K., Walsh J.R., 2006, MNRAS, 451, 925

\refb Habing H.J., 1996, A\&AR, 7, 97

\refb Josselin E., Mauron N., Planesas P., Bachiller R., 2000, A\&A, 362, 255

\refb Jurcsik J., 1993, in IAU Symp 155, R. Weinberger \&\ A.
Acker Eds. (Kluwer Dordrecht), p.399

\refb Mauron M., Huggins P.J., 2006, A\&A, 452, 257

\refb Miranda L,\,F., Torrelles J.\,M., Guerrero M.\,A., 
V\'azquez R., G\'omez Y., 2001, MNRAS, 321, 487

\refb Mitchell D.L, Pollacco D., O'Brien T.J., Lopez J.A., 
Meaburn J., Vaytet N.M.H., 2006, in preparation

\refb Moe M., de Marco O., 2006, ApJ, in press. astro-ph/0606354

\refb Pollacco D.L., Bell S.A., 1997, MNRAS, 284, 32

\refb Roth M., Echevarria J., Tapia M., Carrasco L., Costero R.,
 Rodriguez L.F., 1984, A\&A, 137, L9

\refb Shen Z.-X., Liu X.-W., Danziger I.J, 2004, A\&A, 422, 563 

\refb Soker N., 1998, ApJ, 496, 833

\refb Soker N., Subag E., 2005, AJ, 130, 2717  

\refb Strassmeier K.\,G., Hubl B., Rice J.\,B., 1997, A\&A, 322, 511

\refb van Winckel H., 2003, ARA\&A, 41, 391

\refb Walsh J.\,R., Walton N, 1996, A\&A, 315, 253

\refb Woitke P., 2006, A\&A, submitted (astroph0609392)

\end{document}